\documentclass[12pt,draftclsnofoot,letter,onecolumn]{IEEEtran}

\usepackage{amsmath,amssymb,verbatim,cite,amsopn,graphicx,citesort,url,color}

\renewcommand{\baselinestretch}{1.5}

\newcommand{\SIR}{{\rm{SIR}}}

\newcommand{\Pout}{{\rm{P}_{\rm{co}}}}
\newcommand{\Pint}{{\rm{P}_{\rm{so}}}}
\newcommand{\PintUB}{{\rm{P}_{\rm{so}}^{\rm{UB}}}}
\newcommand{\PintLB}{{\rm{P}_{\rm{so}}^{\rm{LB}}}}

\def\onedot{.,\,}

\def\ie{i.e\onedot}

\def\wrt{w.r.t.\,}

\newcommand{\AuthorOne}{Xiangyun Zhou}
\newcommand{\AuthorTwo}{Radha Krishna Ganti}
\newcommand{\AuthorThree}{Jeffrey G. Andrews}
\newcommand{\AuthorFour}{Are Hj{\o}rungnes}

\newcommand{\ThankTwo}{X. Zhou and A. Hj{\o}rungnes are with UNIK - University Graduate Center, University of Oslo, Kjeller, NO-2027, Norway. (Email: \{xiangyun,arehj\}@unik.no). R. K. Ganti and J. G. Andrews are with the Department of Electrical and Computer Engineering, the University of Texas at Austin, Austin, TX 78712, USA. (Emails: rganti@austin.utexas.edu, jandrews@ece.utexas.edu). This work was supported by the Research Council of Norway through the project 197565/V30, and the DARPA IT-MANET project. Manuscript date: \today.}

\newtheorem{Theorem}{Theorem}
\newtheorem{Lemma}{Lemma}

\newtheorem{Corollary}{Corollary}

\begin{document}

\title{On the Throughput Cost of Physical Layer Security in Decentralized Wireless Networks}

\author{
\authorblockN{\AuthorOne, \textit{Member, IEEE}, \AuthorTwo, \textit{Member, IEEE},\\
\AuthorThree, \textit{Senior Member, IEEE},\\
and \AuthorFour, \textit{Senior Member, IEEE}
\thanks{\ThankTwo}
}}

\maketitle

\begin{abstract}

This paper studies the throughput of large-scale decentralized wireless networks with physical layer security constraints. In particular, we are interested in the question of how much throughput needs to be sacrificed for achieving a certain level of security. We consider random networks where the legitimate nodes and the eavesdroppers are distributed according to independent two-dimensional Poisson point processes. The transmission capacity framework is used to characterize the area spectral efficiency of secure transmissions with constraints on both the quality of service (QoS) and the level of security. This framework illustrates the dependence of the network throughput on key system parameters, such as the densities of legitimate nodes and eavesdroppers, as well as the QoS and security constraints. One important finding is that the throughput cost of achieving a moderate level of security is quite low, while throughput must be significantly sacrificed to realize a highly secure network. We also study the use of a secrecy guard zone, which is shown to give a significant improvement on the throughput of networks with high security requirements.

\end{abstract}


\section{Introduction}

The problem of securing wireless communications at the physical layer has recently drawn considerable attention. In the pioneering works on physical layer security, Wyner~\cite{wyner_75} introduced the wiretap channel for single point-to-point communication, which was extended to broadcast channels with both common and confidential messages by Csisz\'{a}r and K\"{o}rner~\cite{csiszar_78}. Their results showed that perfect secrecy can be achieved if the intended receiver has a stronger channel than the eavesdropper. Recent studies on physical layer security primarily focused on communications involving a small number of nodes with multi-antenna transmission~\cite{li_07,shafiee_07,khisti_09b}, cooperative transmission in relay channels~\cite{lai_08,dong_08} or multiple access channels~\cite{liang_08,tekin_08}. However, few studies have been carried out for large-scale wireless networks. Unlike point-to-point communications, where it is reasonably easy to establish secret keys and have encrypted transmissions, security is more expensive and difficult to achieve in large-scale decentralized networks. Therefore, physical layer security may be important for exchanging secret keys and adding another layer of protection in such networks.

The communication between any pair of nodes in large-scale networks strongly depends on the locations of other nodes and how the nodes interact with each other. When secure communication is required in the presence of eavesdroppers, the locations and channel state information of the eavesdroppers, which are usually unknown, become extra parameters affecting the network throughput. Initial works on network security from an information-theoretic viewpoint mainly considered networks where the legitimate nodes and the eavesdroppers are randomly distributed, and studied the connectivity~\cite{haenggi_08,pinto_10,goel_10,pinto_10b,zhou_11}, coverage~\cite{sarkar_10}, and capacity scaling laws~\cite{koyluoglu_10,liang_09,vasudevan_10}. Specifically, various statistical characterizations of the existence of secure connections were given in~\cite{haenggi_08,pinto_10,goel_10,zhou_11}. Using tools from percolation theory, the existence of a secrecy graph was analyzed in~\cite{haenggi_08,goel_10,pinto_10b}. These connectivity results are concerned with the possibility of having secure communication, while they do not give insight on the network throughput. The authors in~\cite{koyluoglu_10,liang_09,vasudevan_10} derived secrecy capacity scaling laws in static and mobile ad hoc networks, \ie the order-of-growth of the secrecy capacity as the number of nodes increases. Although the scaling laws may provide insights into the information-theoretic performance of large-scale networks, a finer view of throughput is necessary to better understand the impact of key system parameters and transmission protocols, since most of these design choices affect the throughput but not the scaling behaviors~\cite{haenggi_09}.

\subsection{Approach and Contributions}

In this work, we aim to characterize the throughput of secure communications in large wireless networks and to understand how the security requirements affect the network throughput. Our approach uses a metric termed the \textit{transmission capacity}~\cite{weber_05}, which provides the area spectral efficiency (ASE) of decentralized networks with random topology, identical nodes, and a constraint on outage probability. A tutorial on transmission capacity can be found in~\cite{weber_10}, which showed how analytical results can often be derived in simple forms. We extend this capacity framework to study the impact of physical layer security requirements on the network ASE.

The networks considered have both legitimate nodes and eavesdroppers, whose locations follow homogeneous Poisson point processes (PPPs). We define the \textit{secrecy transmission capacity} as the achievable rate of successful transmission of confidential messages per unit area for given constraints on the quality of service (QoS) and the level of security. The QoS constraint is given by the outage probability of the transmission between a legitimate transmitter-receiver pair, while the security constraint is given by the probability of a transmission failing to achieve perfect secrecy. The secrecy transmission capacity shows the dependence of the network ASE on the key system parameters, \ie the densities of legitimate nodes and eavesdroppers, as well as the QoS and security constraints.

To illustrate the use of the general capacity formulation, we derive an accurate closed-form lower bound on the secrecy transmission capacity for Rayleigh fading channels. This simple capacity bound gives a quantitative characterization of the throughput cost of physical layer security. Specifically, the throughput reduction for achieving a moderate level of security is relatively small, while a significant amount of throughput needs to be sacrificed to realize a highly secure network. We also give a condition for the existence of positive secrecy transmission capacity. It turns out that the QoS and security constraints as well as the density of eavesdroppers are crucial in determining the existence of positive secrecy transmission capacity.

In order to minimize the throughput cost of achieving high network security, it is worthwhile to consider transmission protocols that are robust against eavesdropping and can be implemented in a decentralized manner. Since insecure transmission is mainly due to the presence of an eavesdropper close to the transmitter, we consider the use of a secrecy guard zone for networks in which the legitimate transmitters are able to detect the existence of eavesdroppers in their vicinities~\cite{pinto_10,koyluoglu_10}.\footnote{The application of secrecy guard zone is not always possible. It is applicable in the scenarios where, for example, the legitimate transmitters are able to physically inspect their surrounding areas~\cite{pinto_10}.} Transmission of confidential messages take place only if no eavesdroppers are found inside the guard zone of the corresponding transmitter. We consider two transmission protocols when eavesdropper(s) are found inside the guard zone, \ie the transmitter either remains silent or produces artificial noise to help the other transmitters. The secrecy transmission capacity is studied for both protocols. Numerical results show that a significant throughput improvement can be achieved from the use of a guard zone for networks with high security requirements.

The rest of the paper is organized as follows: Section~\ref{sec:SysMod} presents the system model and the secrecy transmission capacity formulation. In Section~\ref{sec:STC}, we obtain analytical results on the secrecy transmission capacity in Rayleigh fading channels. In Section~\ref{sec:SGZ}, we investigate the secrecy guard zone with two different transmission protocols. Numerical results are presented in Section~\ref{sec:NumRes} and concluding remarks in Section~\ref{sec:Conc}. A summary of the notation used in this paper is given in Table~\ref{table:notations}.

\section{System Model And Capacity Formulation} \label{sec:SysMod}

\begin{figure}[!t]
     \begin{minipage}[t]{0.4\textwidth}
        \centering
        \includegraphics[width= 1 \textwidth]{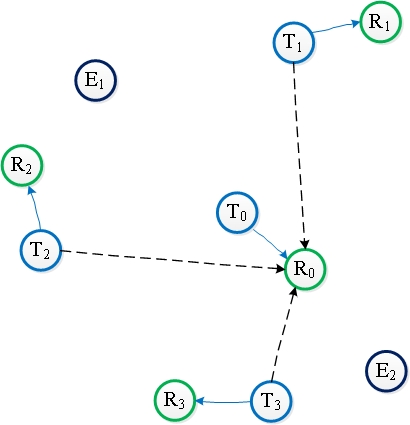}\\
     \vspace{-0mm}
         $(a)$
     \end{minipage}
     \qquad \qquad \qquad \qquad
     \begin{minipage}[t]{0.4\textwidth}
         \centering
         \includegraphics[width= 1 \textwidth]{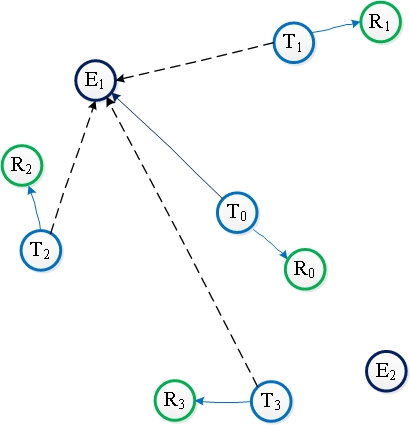}\\
              \vspace{-0mm}
         $(b)$
     \end{minipage}
\vspace{3mm}
    \caption{An example of a part of a network snapshot. Each legitimate transmitter, $\mathrm{T}_i, i=0,1,2,3$, has an intended receiver $\mathrm{R}_i, i=0,1,2,3$, located at a distance $r$. In addition, there are eavesdroppers, $\mathrm{E}_i, i=1,2$, presented in the network. Consider the confidential message transmission from $\mathrm{T}_0$. The intended receiver $\mathrm{R}_0$ as well as each eavesdropper $\mathrm{E}_i, i=1,2$ all try to individually decode the transmitted message. The signal reception at $\mathrm{R}_0$ and (an arbitrary eavesdropper) $\mathrm{E}_1$ are shown in figures (a) and (b), respectively. In both cases, the concurrent transmissions from $\mathrm{T}_1$, $\mathrm{T}_2$ and $\mathrm{T}_3$ act as interference.}
    \label{fig:system}
\end{figure}

We consider an ad hoc network consisting of both legitimate nodes and eavesdroppers over a large two-dimensional space. For each snapshot in time, we have a set of legitimate transmitter locations, denoted by $\Phi_l$.\footnote{For networks employing a slotted Aloha protocol, $\Phi_l$ can be viewed as the locations of the actual transmitters (out of all potential transmitters) in each time slot.} Each transmitter has a unique associated intended receiver. The set of receivers is disjoint with the set of transmitters. In addition, we have a set of eavesdropper locations in each snapshot, denoted by $\Phi_e$. We model $\Phi_l$ and $\Phi_e$ as independent homogeneous PPPs with densities $\lambda_l$ and $\lambda_e$, respectively. This is a suitable model for decentralized networks with nodes having substantial mobility~\cite{weber_10}. Note that the eavesdroppers need to have similar mobility and other behaviors as the legitimate nodes since they can be easily identified otherwise~\cite{liang_09}. Furthermore, we assume that the eavesdroppers do not collude with each other and, hence, must decode the confidential messages individually. An example of a network snapshot is shown in Fig.~\ref{fig:system}.

Consider only one active transmitter that wants to send confidential messages to its intended receiver in the presence of the eavesdroppers. Secure encoding schemes, such as the Wyner code~\cite{wyner_75}, were found in point-to-point systems with the notion of weak secrecy. The notion of strong secrecy and the corresponding encoding scheme were studied in~\cite{subramanian_11}. According to Wyner's encoding scheme, the transmitter chooses two rates, namely, the rate of the transmitted codewords $R_t$ and the rate of the confidential messages $R_s$. The rate difference $R_e = R_t-R_s$ reflects the cost of securing the messages against eavesdropping. If $R_t$ is less than the mutual information between the channel input and output of the legitimate link, the receiver is able to decode the message with an arbitrarily small error. At the same time, if $R_e$ is larger than the mutual information between the channel input and output of every eavesdropper link (\ie links from the transmitter to every eavesdropper), perfect secrecy is achieved as the mutual information between the confidential message and every eavesdropper's received signal approaches zero ratewise. The detailed description of the Wyner code can be found in~\cite{wyner_75,thangaraj_07,tang_09}.

In an ad hoc network with simultaneous transmissions from infinitely many legitimate transmitters, it is difficult to study the mutual information between any pair of nodes. To make the design and analysis mathematically tractable, we assume that the transmitted signal (\ie channel input) has a Gaussian distribution and both the intended receivers and the eavesdroppers treat the interference from concurrent transmissions as noise. In addition, we assume that the network is interference-limited, hence, the receiver noise is negligible. With these assumptions, the mutual information or capacity of either a legitimate link or an eavesdropper link is now determined by the instantaneous signal to interference ratio (SIR). For any given choices of $R_t$ and $R_s$ in Wyner's encoding scheme, the following outage events can result from any transmission~\cite{tang_09}:
\begin{itemize}
 \item \textbf{Connection Outage}: The capacity of the channel from the transmitter to the intended receiver is below the transmission rate $R_t$. Hence, the message cannot be correctly decoded by the intended receiver. The probability of this event happening is referred to as the \textit{connection outage probability}, denoted as $\Pout$.
 \item \textbf{Secrecy Outage}: The capacity of the channel from the transmitter to one or more eavesdroppers is above the rate $R_e$. Hence, the message is not perfectly secure against eavesdropping. The probability of this event happening is referred to as the \textit{secrecy outage probability}, denoted as $\Pint$.
\end{itemize}
The connection outage probability can be regarded as the communication QoS while the secrecy outage probability gives a measure of the security level.

The primary goal of this work is to characterize the throughput of secure transmissions in decentralized wireless networks. Although it is extremely difficult to find the network capacity region, the idea of transmission capacity proposed in~\cite{weber_05} often gives useful insights on the network ASE and the impacts of the key system parameters. Building on the existing transmission capacity framework, we define the \textit{secrecy transmission capacity} as the achievable rate of successful transmission of confidential messages per unit area, for a given connection outage constraint and a given secrecy outage constraint. Mathematically, the secrecy transmission capacity, with a connection outage probability of $\Pout = \sigma$ and a secrecy outage probability of $\Pint = \epsilon$, is defined as
\begin{eqnarray}\label{eq:STCgen0}
        \tau = \bar{R}_s (1-\sigma)\lambda_l,
\end{eqnarray}
where $\bar{R}_s$ is the average rate of confidential messages. In this work, we focus on a simple scenario where the transmit power and the distances to the intended receivers have fixed values which are the same for all transmitters. Therefore, the confidential message rates also take the same values for all transmitters. Denote the distance between the legitimate transmitter-receiver pairs as $r$, the secrecy transmission capacity can be written as
\begin{eqnarray}\label{eq:STCgen}
        \tau(r) = R_s (1-\sigma)\lambda_l.
\end{eqnarray}
The connection outage constraint $\sigma$ determines the value of $R_t$, while the secrecy outage constraint $\epsilon$ determines the value of $R_e$. Therefore, the rate of confidential messages $R_s$ in (\ref{eq:STCgen}), given by $R_t-R_e$, is a function of $\sigma$ and $\epsilon$. With these choices of rates for the Wyner code, the probability that a message transmission can be successfully decoded by the intended receiver is $1-\sigma$, while the probability that a message transmission is perfectly secure against eavesdropping is $1-\epsilon$, under the assumption of treating interference as noise.

If we allow the distance between the legitimate transmitter-receiver pair varies over time and/or space but follows some known distribution $f(r)$, the secrecy transmission capacity is computed as
\begin{eqnarray}
        \tau = \int \tau(r) f(r) \mathrm{d}r.
\end{eqnarray}
In practice, the distribution of $r$ depends on specific scenarios. Hence, we do not consider the variation in $r$ and focus on $\tau(r)$ in (\ref{eq:STCgen}).

\section{Secrecy Transmission Capacity in Rayleigh Fading Channels} \label{sec:STC}

In this section, we derive analytical results on the secrecy transmission capacity for Rayleigh fading channels. We assume that each node has a single antenna for transmission or reception, and the fading channel states are known at the receiver side (including the eavesdroppers) but not at the transmitter side. The derivation of the secrecy transmission capacity involves two main steps: 1) Use the connection outage constraint $\sigma$ to find the value of $R_t$. 2) Use the secrecy outage constraint $\epsilon$ to find the value of $R_e$.

Our analysis is based on an arbitrarily chosen transmitter-receiver pair, which are named the typical transmitter and receiver. For confidential message transmission from the typical transmitter, the other transmitters act as interferers to the typical receiver or any eavesdropper. From Slivnyak's Theorem~\cite{stoyan_96}, the spatial distribution of the interferers, given the location of the typical transmitter, still follows a homogeneous PPP with density $\lambda_l$. By slight abuse of notation (since we have used $\Phi_l$ to denote the set of all transmitter locations), we will also refer to $\Phi_l$ as the set of interferer locations in the rest of this paper.

For the typical receiver, a connection outage occurs if $\log_2(1+\SIR_0) < R_t$, where $\SIR_0$ denotes the SIR at the typical receiver given by
\begin{eqnarray}\label{eq:SIR1}
         \SIR_0 = \frac{S_0 r^{-\alpha}}{\sum_{l\in\Phi_l} S_l |X_l|^{-\alpha}},
\end{eqnarray}
where $S_0$ and $r$ are the channel fading gain and the distance between the typical transmitter and receiver, respectively, $\alpha$ is the path loss exponent, $S_l$ and $|X_l|$ are the channel fading gain and the distance between the interferer (at position) $l$ in $\Phi_l$ and the typical receiver, respectively. We assume $\alpha > 2$ throughout this paper. The fading gains are modeled as independent and identically distributed (i.i.d.) exponential random variables with unit mean.

Define a threshold SIR value for connection outage as
\begin{eqnarray}\label{eq:thrSIR}
         \beta_t = 2^{R_t}-1.
\end{eqnarray}
Hence, the connection outage probability can be written as
\begin{eqnarray}
         \Pout= \mathbb{P}\Big(\SIR_0 < \beta_t\Big)
         =\mathbb{P} \left(\frac{S_0 r^{-\alpha}}{\sum_{l\in\Phi_l} S_l |X_l|^{-\alpha}}<\beta_t\right).
\end{eqnarray}
The summation term $\sum_{l\in\Phi_l} S_l |X_l|^{-\alpha}$ is a shot noise process~\cite{venkataraman_06} in two-dimensional space whose Laplace transform is known in a closed form and was used to compute the connection outage probability in~\cite{baccelli_06} as
\begin{eqnarray}\label{eq:Pout}
         \Pout &=& 1-\exp\left[-\lambda_l\pi r^2\beta_t^{2/\alpha}
         \Gamma\Big(1-\frac{2}{\alpha}\Big)\Gamma\Big(1+\frac{2}{\alpha}\Big)\right].
\end{eqnarray}

With the connection outage constraint given by $\Pout = \sigma$, the transmission rate $R_t$ can be found using (\ref{eq:thrSIR}) and (\ref{eq:Pout}) as
\begin{eqnarray}\label{eq:Rb}
         R_t &=& \log_2\left(1+\Bigg[\frac{\ln\frac{1}{1-\sigma}}{\lambda_l\pi r^2\Gamma\Big(1-\frac{2}{\alpha}\Big)\Gamma\Big(1+\frac{2}{\alpha}\Big)}
         \Bigg]^{\frac{\alpha}{2}} \right).
\end{eqnarray}
It is clear that a lower connection outage probability (\ie a higher QoS) requires a lower $R_t$.

On the other hand, the confidential message transmission is not perfectly secure against the eavesdropper (at position) $e$ in $\Phi_e$ if $\log_2(1+\SIR_e) > R_e$, where $\SIR_e$ denotes the SIR at $e$ given by
\begin{eqnarray}
         \SIR_e = \frac{S_e |X_e|^{-\alpha}}{\sum_{l\in\Phi_l} S_{le} |X_{le}|^{-\alpha}},
\end{eqnarray}
where $S_e$ and $|X_e|$ are the channel fading gain and the distance between the typical transmitter and eavesdropper $e$ in $\Phi_e$, respectively, $S_{le}$ and $|X_{le}|$ are the channel fading gain and the distance between node $l$ in $\Phi_l$ and eavesdropper $e$ in $\Phi_e$, respectively. The fading gains are modeled as i.i.d. exponential random variables with unit mean.

Define a threshold SIR value for secrecy outage as
\begin{eqnarray}\label{eq:thrSIR2}
         \beta_e = 2^{R_e}-1.
\end{eqnarray}
Let $A=\{y \in \Phi_e : \SIR_y > \beta_e\}$, \ie the set of eavesdroppers that can cause secrecy outage. Hence, we can define the following indicator function: $1_A(e)$, which equals 1 when the eavesdropper $e$ is in the set $A$. The secrecy outage probability equals the probability that at least one of the eavesdroppers in $\Phi_e$ causes a secrecy outage, which can be written as
\begin{eqnarray}\label{eq:Pint1st}
         \Pint &=& 1-\mathbb{E}_{\Phi_l}\Big\{\mathbb{E}_{\Phi_e}\Big\{
         \mathbb{E}_{S}\Big\{
         \prod_{e\in\Phi_e}\Big(1-1_A(e)\Big)\Big\}\Big\}\Big\},\nonumber\\
         &=& 1-\mathbb{E}_{\Phi_l}\Bigg\{\mathbb{E}_{\Phi_e}\Bigg\{
         \prod_{e\in\Phi_e}\Bigg(1-\mathbb{P} \Big(\frac{S_e |X_e|^{-\alpha}}{\sum_{l\in\Phi_l} S_{le} |X_{le}|^{-\alpha}} >\beta_e\Big|\Phi_e, \Phi_l \Big)\Bigg)\Bigg\}\Bigg\}.
\end{eqnarray}
where the independence in the fading gains among different eavesdroppers is used to move the expectation over $S=\{S_e, S_{le}\}$ inside the product over $\Phi_e$ in (\ref{eq:Pint1st}). Since it is difficult to express $\Pint$ in a closed form, we look for analytical bounds on the secrecy outage probability. The results are summarized in the following lemma:

\begin{Lemma}\label{Lemma:1}
{\em{The secrecy outage probability is bounded from above by
\begin{eqnarray}\label{eq:PintUB}
         \PintUB = 1-\exp\left[-\frac{\lambda_e}{\lambda_l\beta_e^{2/\alpha}
         \Gamma\Big(1-\frac{2}{\alpha}\Big)\Gamma\Big(1+\frac{2}{\alpha}\Big)}\right],
\end{eqnarray}
and bounded from below by
\begin{eqnarray}\label{eq:PintLB}
         \PintLB = \frac{1}{1+\frac{\lambda_l}{\lambda_e}\beta_e^{2/\alpha}
         \Gamma\Big(1-\frac{2}{\alpha}\Big)\Gamma\Big(1+\frac{2}{\alpha}\Big)}.
\end{eqnarray}}}
\end{Lemma}

{\em{Proof:}} Using the generating functional of the PPP $\Phi_e$~\cite{stoyan_96}, we can express the secrecy outage probability in (\ref{eq:Pint1st}) as
\begin{eqnarray}\label{eq:}
         \Pint = 1-\mathbb{E}_{\Phi_l}\Bigg\{\exp\Bigg[-\lambda_e
         \int_{\mathbb{R}^2} \mathbb{P} \Big(\frac{S_e |X_e|^{-\alpha}}{\sum_{l\in\Phi_l} S_{le} |X_{le}|^{-\alpha}} >\beta_e \Big| \Phi_l\Big)\mathrm{d}e\Bigg]\Bigg\}.
\end{eqnarray}
Jensen's inequality gives an upper bound on $\Pint$
\begin{eqnarray}
         \Pint &\leq& 1-\exp\Bigg[-\lambda_e
         \int_{\mathbb{R}^2}\mathbb{P} \Big(\frac{S_e |X_e|^{-\alpha}}{\sum_{l\in\Phi_l} S_{le} |X_{le}|^{-\alpha}} >\beta_e\Big)\mathrm{d}e\Bigg]\nonumber\\
         &=&1-\exp\Bigg[-2\pi\lambda_e\int_0^\infty
         \exp\left[-\lambda_l\pi r_e^2\beta_e^{2/\alpha}
         \Gamma\Big(1-\frac{2}{\alpha}\Big)\Gamma\Big(1+\frac{2}{\alpha}\Big)\right]
         r_e \mathrm{d}r_e\Bigg],\label{eq:Pintub1}
\end{eqnarray}
where $r_e$ denotes the distance between the typical transmitter and eavesdropper $e$, (\ref{eq:Pintub1}) is arrived in the same way as (\ref{eq:Pout}) followed by changing to polar coordinates. The upper bound in (\ref{eq:PintUB}) is then obtained by directly evaluating the integration in (\ref{eq:Pintub1}).

The lower bound on $\Pint$ is obtained by considering only the eavesdropper nearest to the typical transmitter. Denote the eavesdropper (location) in $\Phi_e$ that is nearest to the typical transmitter as ${e'}$ and denote the distance between ${e'}$ and the typical transmitter as $r_{e'}$. The probability distribution of $r_{e'}$ is given by~\cite{haenggi_05}
\begin{eqnarray}\label{eq:}
        f(r_{e'}) = 2\lambda_e\pi r_{e'} \exp(-\lambda_e \pi r_{e'}^2).
\end{eqnarray}
The secrecy outage probability is bounded from below by the probability that the nearest eavesdropper causes a secrecy outage, \ie
\begin{eqnarray}
         \Pint &\geq& \int_0^\infty
         \mathbb{P} \Big(\frac{S_{e'} r_{e'}^{-\alpha}}{\sum_{l\in\Phi_l} S_{l{e'}} |X_{l{e'}}|^{-\alpha}} >\beta_e\Big)f(r_{e'})\mathrm{d}r_{e'}\nonumber\\
         &=& \int_0^\infty
         \exp\left[-\lambda_l\pi r_{e'}^2\beta_e^{2/\alpha}
         \Gamma\Big(1-\frac{2}{\alpha}\Big)\Gamma\Big(1+\frac{2}{\alpha}\Big)\right]
         2\lambda_e\pi r_{e'} \exp(-\lambda_e \pi r_{e'}^2)\mathrm{d}r_{e'}.\label{eq:Pintlb1}
\end{eqnarray}
The lower bound in (\ref{eq:PintLB}) is then obtained by directly evaluating the integration in (\ref{eq:Pintlb1}).\hfill$\blacksquare$

Note that the authors in~\cite{ganti_07} used the same bounding techniques to derive analytical bounds on the probability of connectivity in a different network scenario and numerically studied the accuracy of the derived bounds. From the numerical illustration in~\cite[Fig.~5]{ganti_07}, we know that the upper bound $\PintUB$ in (\ref{eq:PintUB}) gives accurate approximation of the exact secrecy outage probability over the entire range of $\Pint\in[0,1]$, while the lower bound $\PintLB$ in (\ref{eq:PintLB}) is usually very different from the exact value of $\Pint$. Moreover, both $\PintUB$ and $\PintLB$ are asymptotically tight in the low probability regime. To see this, we consider $\PintUB \approx 0$ and $\PintLB \approx 0$, in which case the bounds in (\ref{eq:PintUB}) and (\ref{eq:PintLB}) can be approximated by
\begin{eqnarray}\label{eq:lowPintApp}
         \PintUB \approx \frac{\lambda_e}{\lambda_l\beta_e^{2/\alpha}         \Gamma\Big(1-\frac{2}{\alpha}\Big)\Gamma\Big(1+\frac{2}{\alpha}\Big)} \approx \PintLB.
\end{eqnarray}
Hence, both $\PintUB$ and $\PintLB$ approach the exact value of $\Pint$ in the low probability regime.

Recall that the goal here is to determine the value of $R_e$ from the secrecy outage constraint of $\Pint = \epsilon$. Using the upper bound on the secrecy outage probability in (\ref{eq:PintUB}), the value of $R_e$ that guarantees the required security level can be found as
\begin{eqnarray}\label{eq:Re}
         R_e = \log_2\left(1+\Big[\frac{\lambda_l}{\lambda_e}
         \Gamma\Big(1-\frac{2}{\alpha}\Big)\Gamma\Big(1+\frac{2}{\alpha}\Big)\ln\frac{1}{1-\epsilon}
         \Big]^{-\frac{\alpha}{2}} \right).
\end{eqnarray}
It is clear that a lower secrecy outage probability (\ie a higher security level) requires a higher $R_e$.


Having $R_t$ in (\ref{eq:Rb}) and $R_e$ in (\ref{eq:Re}), a lower bound on the secrecy transmission capacity is obtained as $\tau^{\rm{LB}}(r) = (R_t-R_e) (1-\sigma)\lambda_l$. Its expression is presented in the following theorem:

\begin{Theorem}\label{Theorem:1}
{\em{A lower bound on the secrecy transmission capacity with a connection outage constraint of $\sigma$ and a secrecy outage constraint of $\epsilon$ is given by
\begin{eqnarray}\label{eq:STCR}
        \tau^{\rm{LB}}(r) = (1-\sigma)\lambda_l \log_2\left(\frac{1+\Big[\frac{\ln\frac{1}{1-\sigma}}{\lambda_l\pi r^2\Gamma(1-\frac{2}{\alpha})\Gamma(1+\frac{2}{\alpha})}
         \Big]^{\frac{\alpha}{2}}}
        {1+\Big[\frac{\lambda_l}{\lambda_e}
         \Gamma\Big(1-\frac{2}{\alpha}\Big)\Gamma\Big(1+\frac{2}{\alpha}\Big)\ln\frac{1}{1-\epsilon}
         \Big]^{-\frac{\alpha}{2}}}
        \right).
\end{eqnarray}}}
\end{Theorem}

From our discussion on the accuracy of $\PintUB$, we know that the lower bound on the secrecy transmission capacity in (\ref{eq:STCR}) is generally accurate for any values of $\sigma$ and $\epsilon$, and is asymptotically tight as $\epsilon \rightarrow 0$. Therefore, we will for simplicity refer to $\tau^{\rm{LB}}(r)$ in (\ref{eq:STCR}) as the secrecy transmission capacity in the rest of this paper. It is clear from (\ref{eq:STCR}) that $\tau^{\rm{LB}}(r)$ reduces as $\epsilon$ decreases. The reduction in $\tau^{\rm{LB}}(r)$ as $\epsilon$ decreases can be viewed as the throughput cost of improving physical layer security.

In practical network design, the connection outage constraint and the spatial transmission intensity\footnote{In networks employing an Aloha protocol, the spatial transmission intensity equals the density of potential transmitters multiplied by the probability of transmission. In this case, the system designer may control the probability of transmission to vary the spatial transmission intensity.} may be under the control of the system designer. The derived closed-form characterization of the secrecy transmission capacity allows the designer to optimize these system parameters to maximize the throughput of secure transmissions with a target security level.

\subsection{Existence of Positive Secrecy Transmission Capacity}\label{sec:PosSTC}

A fundamental question to ask is the condition under which positive secrecy transmission capacity exists. From the expression in (\ref{eq:STCR}), one can find this condition by solving $\tau^{\rm{LB}}(r)>0$.

\begin{Corollary}\label{Corollary:1}
{\em{The condition for positive secrecy transmission capacity is given by
\begin{eqnarray}\label{eq:Req}
        \ln\frac{1}{1-\sigma}\ln\frac{1}{1-\epsilon} &>& \pi r^2 \lambda_e.
\end{eqnarray}}}
\end{Corollary}
In other words, positive secrecy transmission capacity is achieved if the average number of eavesdroppers within a distance $r$ from the transmitter (\ie having shorter distances than the intended receiver) is less than $\ln\frac{1}{1-\sigma}\ln\frac{1}{1-\epsilon}$.

\textit{Remark 1}: The condition in (\ref{eq:Req}) clearly gives a trade-off between the QoS and the security level of a network: The QoS needs to be compromised (\ie allowing a larger value of $\sigma$) in order to achieve a higher security level (\ie a smaller value of $\epsilon$). Therefore, a moderate connection outage probability is usually desirable for highly secure networks. Furthermore, the feasible range of $\sigma$ can be found from (\ref{eq:Req}) as
\begin{eqnarray}\label{eq:OutRange}
        \sigma \in \left(1-\exp\left[-\frac{\pi r^2\lambda_e}{\ln\frac{1}{1-\epsilon}}\right],1\right).
\end{eqnarray}

\textit{Remark 2}: The condition in (\ref{eq:Req}) does not depend on the spatial transmission intensity $\lambda_l$. That is to say, positive secrecy transmission capacity cannot be achieved simply by bringing in additional legitimate users or deactivating existing legitimate users, if the connection outage and secrecy outage performances of the network do not meet the condition in (\ref{eq:Req}). Once this condition is met and the network is operating with some positive secrecy transmission capacity, there exists an optimal value of $\lambda_l$ which can be found numerically using (\ref{eq:STCR}). To obtain some analytical insights into the optimal $\lambda_l$, we consider the low secrecy transmission capacity regime by letting $\ln\frac{1}{1-\sigma}\ln\frac{1}{1-\epsilon} \approx \pi r^2 \lambda_e$ which implies $\tau^{\rm{LB}}(r) \approx 0$. We rewrite (\ref{eq:STCR}) as
\begin{eqnarray}\label{eq:STCRlow}
        \tau^{\rm{LB}}(r) &=& (1-\sigma)\lambda_l \log_2\left(1+
        \frac{\Big[\frac{\ln\frac{1}{1-\sigma}}{\lambda_l\pi r^2\Gamma(1-\frac{2}{\alpha})\Gamma(1+\frac{2}{\alpha})}
         \Big]^{\frac{\alpha}{2}}-\Big[\frac{\lambda_l}{\lambda_e}
         \Gamma\Big(1-\frac{2}{\alpha}\Big)\Gamma\Big(1+\frac{2}{\alpha}\Big)\ln\frac{1}{1-\epsilon}
         \Big]^{-\frac{\alpha}{2}}}
        {1+\Big[\frac{\lambda_l}{\lambda_e}
         \Gamma\Big(1-\frac{2}{\alpha}\Big)\Gamma\Big(1+\frac{2}{\alpha}\Big)\ln\frac{1}{1-\epsilon}
         \Big]^{-\frac{\alpha}{2}}}
        \right)\nonumber\\
        &\approx& \frac{(1-\sigma)\lambda_l }{\ln 2}\,
        \frac{\Big[\frac{\ln\frac{1}{1-\sigma}}{\lambda_l\pi r^2\Gamma(1-\frac{2}{\alpha})\Gamma(1+\frac{2}{\alpha})}
         \Big]^{\frac{\alpha}{2}}-\Big[\frac{\lambda_l}{\lambda_e}
         \Gamma\Big(1-\frac{2}{\alpha}\Big)\Gamma\Big(1+\frac{2}{\alpha}\Big)\ln\frac{1}{1-\epsilon}
         \Big]^{-\frac{\alpha}{2}}}
        {1+\Big[\frac{\lambda_l}{\lambda_e}
         \Gamma\Big(1-\frac{2}{\alpha}\Big)\Gamma\Big(1+\frac{2}{\alpha}\Big)\ln\frac{1}{1-\epsilon}
         \Big]^{-\frac{\alpha}{2}}}\nonumber\\
         &=& \frac{1-\sigma}{\ln 2}
         \Big[\Gamma\Big(1-\frac{2}{\alpha}\Big)\Gamma\Big(1+\frac{2}{\alpha}\Big)\Big]^{-\frac{\alpha}{2}}
         \left[\left(\frac{\ln\frac{1}{1-\sigma}}{\pi r^2}\right)^{\frac{\alpha}{2}}
         -\left(\frac{\ln\frac{1}{1-\epsilon}}{\lambda_e}\right)^{-\frac{\alpha}{2}}\right]\nonumber\\
         && \times \frac{1}
         {\lambda_l^{\frac{\alpha}{2}-1}+\Big[\Gamma\Big(1-\frac{2}{\alpha}\Big)\Gamma\Big(1+\frac{2}{\alpha}\Big)
         \frac{\ln\frac{1}{1-\epsilon}}{\lambda_e}\Big]^{-\frac{\alpha}{2}}\lambda_l^{-1}}.
\end{eqnarray}
Assuming $\tau^{\rm{LB}}(r)$ in (\ref{eq:STCRlow}) is positive, the optimal value of $\lambda_l$ that maximizes $\tau^{\rm{LB}}(r)$ is given by
\begin{eqnarray}\label{eq:OptDenSol}
        \lambda_l^{\rm{opt}} = \Big(\frac{2}{\alpha-2}\Big)^{\frac{2}{\alpha}}
        \frac{\lambda_e}{\Gamma\Big(1-\frac{2}{\alpha}\Big)\Gamma\Big(1+\frac{2}{\alpha}\Big)\ln\frac{1}{1-\epsilon}}.
\end{eqnarray}
From (\ref{eq:OptDenSol}), we see that the optimal spatial transmission intensity increases as the required security level increases (\ie as $\epsilon$ decreases). In addition, the optimal spatial transmission intensity is usually much higher than the density of eavesdroppers for highly secure networks. For example, $\lambda_l^{\rm{opt}}/\lambda_e \approx 63 $ for $\epsilon = 0.01$ and $\alpha = 4$. Although these observations are made in the regime of arbitrarily low secrecy transmission capacity, as we will see in Section~\ref{sec:NumRes}, they are also valid for more general scenarios.

\subsection{Optimal Connection Outage Probability in Sparse Networks}\label{sec:OptCon}

When the system designer has control over the connection outage constraint, the expression of secrecy transmission capacity in (\ref{eq:STCR}) can be used to numerically find the value of $\sigma$ that maximizes $\tau^{\rm{LB}}(r)$. Here, we present a closed-form solution of the optimal connection outage probability in sparse networks, \ie $\lambda_l \pi r^2 \ll 1$. Note that our analysis is based on the assumption of interference-limited networks, which is valid if the transmit power of the legitimate users is sufficiently high such that the receiver noise is much weaker than the aggregate interference.

From the discussion in Subsection~\ref{sec:PosSTC}, we know that the value of $\sigma$ should not be chosen very close to 0. When the network is sparse, \ie $\lambda_l \pi r^2 \ll 1$, the secrecy transmission capacity in (\ref{eq:STCR}) can be approximated as
\begin{eqnarray}
        \tau^{\rm{LB}}(r) &\approx& (1-\sigma)\lambda_l \log_2\left(\frac{\Big[\frac{\ln\frac{1}{1-\sigma}}{\lambda_l\pi r^2\Gamma(1-\frac{2}{\alpha})\Gamma(1+\frac{2}{\alpha})}
         \Big]^{\frac{\alpha}{2}}}
        {1+\Big[\frac{\lambda_l}{\lambda_e}
         \Gamma\Big(1-\frac{2}{\alpha}\Big)\Gamma\Big(1+\frac{2}{\alpha}\Big)\ln\frac{1}{1-\epsilon}
         \Big]^{-\frac{\alpha}{2}}}
        \right)\label{eq:STCRS1}\\
        &=&(1-\sigma)\lambda_l \log_2\left( \Big[\kappa \ln\frac{1}{1-\sigma}\Big]^{\frac{\alpha}{2}}\right),\label{eq:STCRS2}
\end{eqnarray}
where we have assumed in (\ref{eq:STCRS1}) that the path loss exponent $\alpha$ is not close to 2 (which happens in most outdoor scenarios) and the connection outage probability is not close to 0, and
\begin{eqnarray}\label{eq:}
         \kappa  =  \frac{\Bigg(1+\Big[\frac{\lambda_l}{\lambda_e}
         \Gamma\Big(1-\frac{2}{\alpha}\Big)\Gamma\Big(1+\frac{2}{\alpha}\Big)\ln\frac{1}{1-\epsilon}
         \Big]^{-\frac{\alpha}{2}}\Bigg)^{-\frac{2}{\alpha}}}{\lambda_l\pi r^2\Gamma\Big(1-\frac{2}{\alpha}\Big)\Gamma\Big(1+\frac{2}{\alpha}\Big)}.
\end{eqnarray}

The optimal connection outage probability that maximizes the secrecy transmission capacity in (\ref{eq:STCRS2}) is given by
\begin{eqnarray}\label{eq:OptSig}
         \sigma^{\rm{opt}} = 1-\frac{1}{\exp\Big[\frac{1}{\mathrm{W}_0(\kappa)}\Big]},
\end{eqnarray}
where $\mathrm{W}_0(\cdot)$ is the real-valued principal branch of Lambert's W function. This result is obtained by directly taking the derivative of $\tau^{\rm{LB}}(r)$ in (\ref{eq:STCRS2}) \wrt $\sigma$ and solving for the root. Furthermore, one can show that the optimal connection outage probability increases when a higher security level (\ie a lower $\epsilon$) is required.

\section{Secrecy Guard Zone}\label{sec:SGZ}

In this section, we consider simple protocols for improving the secrecy transmission capacity. We assume that the legitimate transmitters are able to detect the existence of eavesdroppers within a finite range. We model this range as a disk with radius $D$ centered at each transmitter and call it the secrecy guard zone. Transmission of confidential messages only happens when there is no eavesdropper inside the secrecy guard zone. As we are concerned with decentralized networks, it is assumed that each transmitter individually decides whether or not to transmit based on the existence of eavesdroppers inside its own guard zone.

The general idea of guard zone is not new and has been applied in wireless ad hoc networks without or with security considerations: In~\cite{hasan_07}, the authors used a guard zone around each receiver such that the receiver is active when there is no interferers inside the guard zone. The authors in~\cite{pinto_10} and~\cite{koyluoglu_10} studied the secure connectivity and secrecy capacity scaling law of ad hoc networks in the presence of eavesdroppers, respectively, and applied a secrecy guard zone around each legitimate node. In this work, we apply a secrecy guard zone around each legitimate transmitter and consider the following two transmission protocols:
\begin{enumerate}
  \item \textbf{Non-Cooperative Transmitters}: The transmitter remains silent when eavesdropper(s) are found inside its secrecy guard zone.
  \item \textbf{Cooperative Transmitters}: The transmitter produces artificial noise when eavesdropper(s) are found inside its secrecy guard zone.
\end{enumerate}

\begin{figure}[!t]
\centering\vspace{-0mm}
\includegraphics[width=0.55\columnwidth]{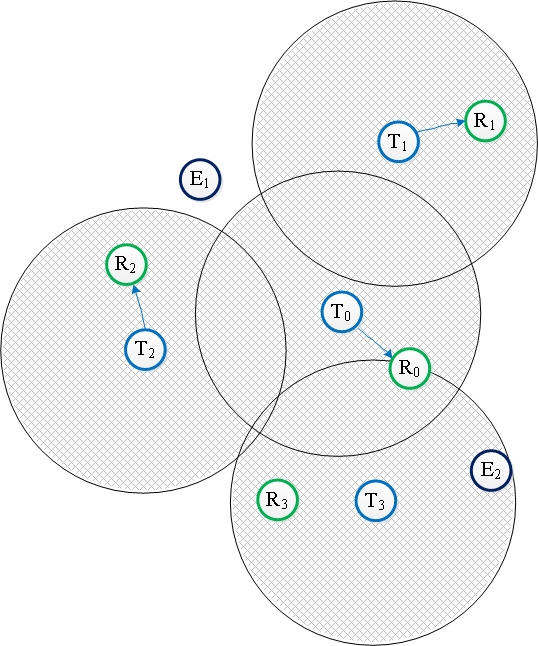}
\vspace{-0mm} \caption{An example of a part of a network snapshot with a secrecy guard zone around each transmitter. Transmitters $\mathrm{T}_0$, $\mathrm{T}_1$, and $\mathrm{T}_2$ do not find any eavesdropper inside their individual guard zone, and hence, transmit confidential messages to their intended receivers. However, transmitter $\mathrm{T}_3$ detects an eavesdropper, $\mathrm{E}_2$, insides its guard zone. If the non-cooperative protocol is used, $\mathrm{T}_3$ remains silent. If the cooperative protocol is used, $\mathrm{T}_3$ transmits artificial noise.} \label{fig:system3}
\end{figure}

The idea of using artificial noise for secrecy was first proposed for multi-antenna transmissions in~\cite{goel_08}, which is also related to the idea of cooperative jamming studied in~\cite{lai_08,tekin_08,ekrem_08,vasudevan_10,tang_11}. An example of a network snapshot with secrecy guard zones is shown in Fig.~\ref{fig:system3}. In the following, we study the secrecy transmission capacity with each transmission protocol.

\subsection{Secrecy Guard Zone with Non-Cooperative Transmitters}\label{sec:SGZnon}

The set of actual transmitter locations, denoted as $\Phi_{l'}$, has density of
\begin{eqnarray}\label{eq:ActTxDen}
         \lambda'_l = \lambda_l\exp[-\pi \lambda_e D^2],
\end{eqnarray}
where the exponential term in (\ref{eq:ActTxDen}) is the probability of no eavesdropper located inside the secrecy guard zone of an arbitrary transmitter. With the secrecy guard zone, the distribution of the actual transmitters does not still follow a homogeneous PPP. The non-homogeneous nature resulted from the introduction of guard zone was discussed in~\cite{hasan_07}. In particular, the authors in~\cite{hasan_07} applied standard Poisson tests to show that the distribution of the actual transmitters can still be well-approximated by a homogeneous PPP outside $\mathcal{B}(b,D)$ from the viewpoint of a receiver at location $b$, where the notation $\mathcal{B}(b,D)$ stands for a disk of radius $D$ centered at $b$. Based on this result, we apply the following two approximations: From the viewpoint of eavesdropper $e$, the actual transmitter $\Phi_{l'}$ follows a homogeneous PPP with density $\lambda'_l$ outside $\mathcal{B}(e,D)$. From the viewpoint of any legitimate receiver, the actual transmitter $\Phi_{l'}$ follows a homogeneous PPP with density $\lambda'_l$.\footnote{Note that the second approximation usually underestimates the interference power at the typical receiver, since the potential interferers near the typical (active) transmitter is more likely to be active than the ones far away from the typical transmitter. However, this underestimation is marginal as long as $\lambda'_l$ is reasonably close to $\lambda_l$, such as the scenarios to be considered in Fig.~\ref{fig:STCvsSecZone}.}

For the typical receiver, the connection outage\footnote{The connection outage event is defined as the \textit{transmitted} message being undecodable at the intended receiver. Hence, it does not include the event of no transmission due to the existence of eavesdropper(s) inside the guard zone. A similar note applies to the secrecy outage event. The effect of transmission probability on the secrecy transmission capacity is incorporated in $\lambda'_l$.} probability is given by
\begin{eqnarray}
         \Pout &=& 1-\exp\left[-\lambda'_l\pi r^2\beta_t^{2/\alpha}
         \Gamma\Big(1-\frac{2}{\alpha}\Big)\Gamma\Big(1+\frac{2}{\alpha}\Big)\right].
\end{eqnarray}
With the connection outage constraint $\Pout = \sigma$ and $\beta_t = 2^{R_t}-1$, the transmission rate $R_t$ is found as
\begin{eqnarray}
         R_t &=& \log_2\left(1+\Bigg[\frac{\ln\frac{1}{1-\sigma}}{\lambda'_l\pi r^2\Gamma\Big(1-\frac{2}{\alpha}\Big)\Gamma\Big(1+\frac{2}{\alpha}\Big)}
         \Bigg]^{\frac{\alpha}{2}} \right).
\end{eqnarray}

From the viewpoint of the typical transmitter located at the origin $o$, the eavesdroppers $\Phi_e$ follows a homogeneous PPP with density $\lambda_e$ outside $\mathcal{B}(o,D)$. Similar to the proof of Lemma~1, an upper bound on the secrecy outage probability is obtained by using the generating functional of $\Phi_e$ and applying Jensen's inequality as
\begin{eqnarray}\label{eq:PintGZ1}
         \PintUB &=& 1-\exp\left[-\lambda_e
         \int_{\mathbb{R}^2\backslash \mathcal{B}(o,D)} \mathbb{P} \Big(\frac{S_e |X_e|^{-\alpha}}{\sum_{l\in\Phi_{l'}} S_{le} |X_{le}|^{-\alpha}} >\beta_e\Big)\mathrm{d}e\right]\nonumber\\
         &=& 1-\exp\left[-\lambda_e
         \int_{\mathbb{R}^2\backslash \mathcal{B}(o,D)} \mathbb{E}_{Z}\Big\{ \exp[-\beta_e|X_e|^{\alpha}Z]
         \Big\}\mathrm{d}e\right]\nonumber\\
         &=&1-\exp\left[-2\pi\lambda_e
         \int_{D}^\infty \mathcal{L}_{Z}(\beta_e r_e^{\alpha})r_e
         \mathrm{d}r_e\right],
\end{eqnarray}
where $Z= \sum_{l\in\Phi_{l'}} S_{le} |X_{le}|^{-\alpha}$ is the sum of interference power at eavesdropper $e$ and $\mathcal{L}_{Z}(\cdot)$ denotes the Laplace transform of $Z$. Note that we have assumed that $\Phi_{l'}$ is a homogeneous PPP outside $\mathcal{B}(e,D)$ from the viewpoint of eavesdropper $e$. Hence, $\mathcal{L}_{Z}(\cdot)$ is given by~\cite{venkataraman_06}
\begin{eqnarray}\label{eq:LT0}
         \mathcal{L}_{Z}(x) &= \exp\Bigg[
         \pi \lambda'_l \Bigg(
         &\!\!\! D^2 \mathbb{E}_{S}\Big\{1-\exp[-x S D^{-\alpha}]\Big\}
         -x^{2/\alpha}\mathbb{E}_{S}\Big\{S^{2/\alpha}\Big\}
         \Gamma\Big(1-\frac{2}{\alpha}\Big)\nonumber\\
         &&\!\!\!+x^{2/\alpha}\mathbb{E}_{S}\Big\{S^{2/\alpha}\Gamma\Big(1-\frac{2}{\alpha},x S D^{-\alpha}\Big)\Big\}  \Bigg)  \Bigg],
\end{eqnarray}
where $S$ is an exponentially distributed random variable with unit mean and $\Gamma(\cdot,\cdot)$ is the upper incomplete Gamma function. Using integral identities from~\cite{gradshteyn_07} to evaluate the expectations in (\ref{eq:LT0}), we have
\begin{eqnarray}\label{eq:LT}
         \mathcal{L}_{Z}(x) &= \exp\Bigg[
         \pi \lambda'_l \Bigg(
         &\!\!\!  \frac{x D^{2-\alpha}}{1+x D^{-\alpha}}
         -x^{2/\alpha}\Gamma\Big(1+\frac{2}{\alpha}\Big)
         \Gamma\Big(1-\frac{2}{\alpha}\Big)\nonumber\\
         &&\!\!\!+\frac{x D^{2-\alpha}}{(1-\frac{2}{\alpha})(1+x D^{-\alpha})^2}\, _2\mathrm{F}_1\Big(1,2;2+\frac{2}{\alpha};\frac{1}{1+x D^{-\alpha}}\Big) \Bigg)  \Bigg],
\end{eqnarray}
where $_2\mathrm{F}_1(\cdot)$ is the Gauss hypergeometric function.

Substituting (\ref{eq:LT}) into (\ref{eq:PintGZ1}), $\PintUB$ is expressed in an integral form, hence, $R_e$ can be solved numerically with the secrecy outage constraint $\PintUB = \epsilon$ and $\beta_e = 2^{R_e}-1$. A lower bound on the secrecy transmission capacity is found as
\begin{eqnarray}\label{eq:STCRG0}
         \tau^{\rm{LB}}(r) = (R_t-R_e)(1-\sigma)\lambda'_l.
\end{eqnarray}

\subsection{Secrecy Guard Zone with Cooperative Transmitters}\label{sec:SGZcop}

In this scenario, the legitimate transmitters cooperative with each other. When eavesdropper(s) are found inside the secrecy guard zone, the transmitter produces artificial noise to help masking the confidential message transmissions from others. The artificial noise is assumed to be statistically identical to the confidential messages and hence, it cannot be distinguished from message transmissions by the eavesdroppers. It is noted that this cooperative protocol is entirely distributive as it does not require any coordination between the legitimate transmitters.

For any legitimate receiver or eavesdropper, the set of interferers is still $\Phi_l$ with density $\lambda_l$. On the other hand, the set of actual transmitters is $\Phi_{l'}$ with density $\lambda'_l$ given by
\begin{eqnarray}\label{eq:}
         \lambda'_l = \lambda_l\exp[-\pi \lambda_e D^2].
\end{eqnarray}

Since the interferers remain the same as if no secrecy guard zone is applied, the connection outage probability $\Pout$ and the transmission rate $R_t$ are still given by (\ref{eq:Pout}) and (\ref{eq:Rb}), respectively.

From the viewpoint of the typical transmitter located at the origin $o$, the eavesdroppers $\Phi_e$ follows a homogeneous PPP with density $\lambda_e$ outside $\mathcal{B}(o,D)$.  Again, an upper bound on the secrecy outage probability is found using the generating functional of $\Phi_e$ and applying Jensen's inequality as
\begin{eqnarray}
         \PintUB &=& 1-\exp\left[-\lambda_e
         \int_{\mathbb{R}^2\backslash \mathcal{B}(o,D)} \mathbb{P} \Big(\frac{S_e |X_e|^{-\alpha}}{\sum_{l\in\Phi_l} S_{le} |X_{le}|^{-\alpha}} >\beta_e\Big)\mathrm{d}e\right]\nonumber\\
         &=&1-\exp\left[-\frac{\lambda_e\exp\Big[
         -\lambda_l\pi\beta_e^{2/\alpha}
         \Gamma\Big(1-\frac{2}{\alpha}\Big)\Gamma\Big(1+\frac{2}{\alpha}\Big)D^2\Big]}
         {\lambda_l\beta_e^{2/\alpha}
         \Gamma\Big(1-\frac{2}{\alpha}\Big)\Gamma\Big(1+\frac{2}{\alpha}\Big)}\right].
\end{eqnarray}
With the secrecy outage constraint of $\PintUB = \epsilon$, we find $R_e$ as
\begin{eqnarray}
         R_e = \log_2\left(1+\left[\frac{\mathrm{W}_0\Big(
         \lambda_e\pi D^2 \Big[\ln\frac{1}{1-\epsilon}\Big]^{-1}\Big)}
         {\lambda_l\pi D^2\Gamma\Big(1-\frac{2}{\alpha}\Big)\Gamma\Big(1+\frac{2}{\alpha}\Big)}
         \right]^{\frac{\alpha}{2}} \right).
\end{eqnarray}

\begin{Theorem}\label{Theorem:2}
{\em{A lower bound on the secrecy transmission capacity for networks having cooperative transmitters with secrecy guard zones is given by
\begin{eqnarray}\label{eq:STCRG}
        \tau^{\rm{LB}}(r) &=& (R_t-R_e) (1-\sigma)\lambda'_l\nonumber\\
        &=& (1-\sigma)\lambda_l\exp[-\pi \lambda_e D^2] \log_2\left(
        \frac{1+\left[\frac{\ln\frac{1}{1-\sigma}}{\lambda_l\pi r^2\Gamma(1-\frac{2}{\alpha})\Gamma(1+\frac{2}{\alpha})}
         \right]^{\frac{\alpha}{2}}}
        {1+\Big[\frac{\mathrm{W}_0\left(
         \lambda_e\pi D^2 [\ln\frac{1}{1-\epsilon}]^{-1}\right)}
         {\lambda_l\pi D^2\Gamma(1-\frac{2}{\alpha})\Gamma(1+\frac{2}{\alpha})}
         \Big]^{\frac{\alpha}{2}}}
        \right).
\end{eqnarray}}}
\end{Theorem}

From the closed-form expression of $\tau^{\rm{LB}}(r)$ in (\ref{eq:STCRG}), we can derive the condition for positive secrecy transmission capacity by solving $\tau^{\rm{LB}}(r)>0$.

\begin{Corollary}\label{Corollary:2}
{\em{The condition for positive secrecy transmission capacity is given by
\begin{eqnarray}\label{eq:ReqG}
\Big[\frac{1}{1-\sigma}\Big]^{(\frac{D}{r})^2}
         \ln\frac{1}{1-\sigma}\ln\frac{1}{1-\epsilon}
         > \pi r^2 \lambda_e.
\end{eqnarray}}}
\end{Corollary}

When the system designer has control over the connection outage constraint and the guard zone size, positive secrecy transmission capacity can be achieved by carefully choosing the values of $\sigma$ and/or $D$ to meet the condition in (\ref{eq:ReqG}). Similar to the networks without guard zones, the density of the legitimate transmitters has no say for the existence of positive secrecy transmission capacity for fixed connection outage probability and guard zone size. Comparing (\ref{eq:Req}) with (\ref{eq:ReqG}), we see that the improvement from having the secrecy guard zone is given by the factor of $[\frac{1}{1-\sigma}]^{(D/r)^2}$. For a fixed $r$, the minimum required value of $D$ increases as $\sigma$ or $\epsilon$ decreases (below certain threshold value). Hence, one may also expect that the optimal value of $D$, which maximizes the secrecy transmission capacity in (\ref{eq:STCRG}), increases as $\sigma$ or $\epsilon$ decreases.

To further improve the network throughput, the secrecy guard zone could in the future be used in combination with other types of guard zone, such as carrier sense multiple access (CSMA) at the transmitters~\cite{kleinrock_75} and the interference guard zone at the receivers~\cite{hasan_07}.

\section{Numerical Results and Discussion}\label{sec:NumRes}

\begin{figure}[!t]
\centering\vspace{-0mm}
\includegraphics[width=0.85\columnwidth]{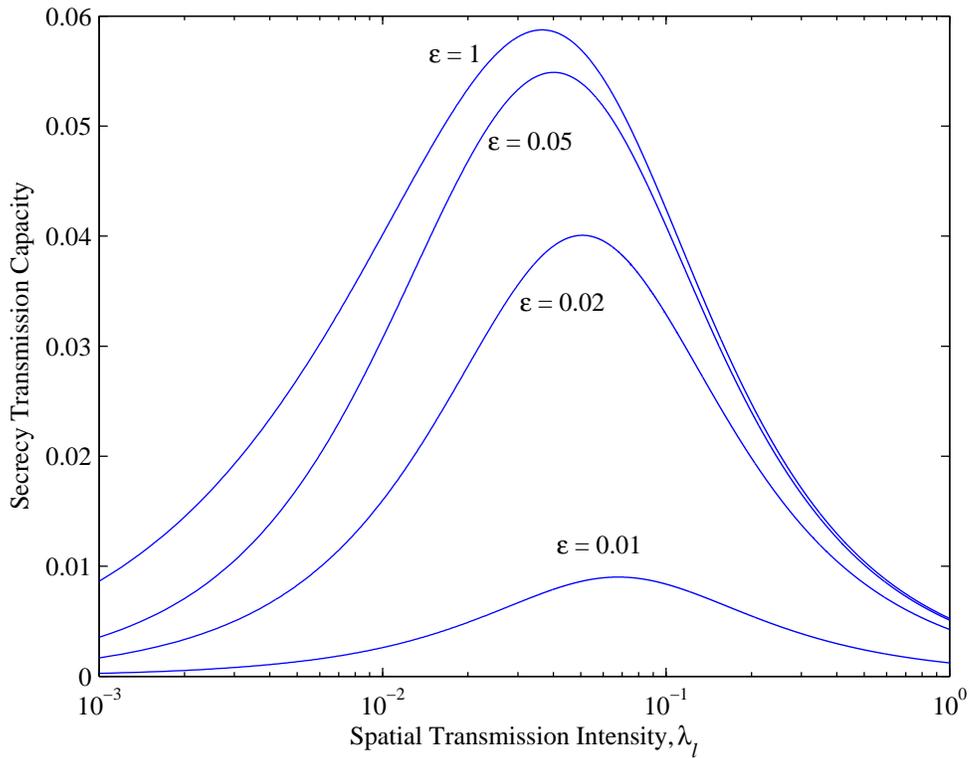}
\vspace{-0mm} \caption{The secrecy transmission capacity $\tau^{\rm{LB}}(r)$ in (\ref{eq:STCR}) versus the density of legitimate transmitters $\lambda_l$. Results are shown for networks with different secrecy outage constraints, \ie $\epsilon=0.01, 0.02, 0.05$, as well as no secrecy constraint, \ie $\epsilon=1$. The other system parameters are $r=1$, $\alpha=4$, $\sigma=0.3$, and $\lambda_e=0.001$.} \label{fig:STCvsLambda}
\end{figure}

In this section, we present numerical results on the secrecy transmission capacity. We first show the interplay of different system parameters and their effects on the secrecy transmission capacity for networks without secrecy guard zones. Fig.~\ref{fig:STCvsLambda} shows the secrecy transmission capacity $\tau^{\rm{LB}}(r)$ in (\ref{eq:STCR}) versus the spatial transmission intensity $\lambda_l$ with different security requirements. Comparing between the four curves, we see that the gap in $\tau^{\rm{LB}}(r)$ between $\epsilon=1$ and $\epsilon=0.05$ is relatively small over a wide range of $\lambda_l$. This suggests that the throughput cost of achieving a moderate security requirement is relatively low. On the other hand, $\tau^{\rm{LB}}(r)$ drops dramatically as $\epsilon$ decreases towards 0. For example, there is a 84\% reduction in $\tau^{\rm{LB}}(r)$ for improving the security level from $\epsilon=0.02$ to $\epsilon=0.01$ at $\lambda_l=0.01$. This reflects a significant increase in the throughput cost of achieving highly secure networks.

For each curve in Fig.~\ref{fig:STCvsLambda}, we see that the optimal value of $\lambda_l$ is generally much larger than $\lambda_e$. This suggests that it is desirable to have a significantly larger number of legitimate nodes than the number of eavesdroppers in the network, which creates a high level of interference to mask the confidential message transmissions against eavesdropping. Furthermore, the optimal value of $\lambda_l$ increases as $\epsilon$ decreases. For example,
the optimal $\lambda_l$ is 0.04 for $\epsilon=0.05$, while it increases to 0.051 for $\epsilon=0.02$ and to 0.068 for $\epsilon=0.01$. Note that the optimal value of $\lambda_l$ computed from (\ref{eq:OptDenSol}) is 0.063 for $\epsilon=0.01$, which is very close to the numerical result.

\begin{figure}[!t]
\centering\vspace{-0mm}
\includegraphics[width=0.85\columnwidth]{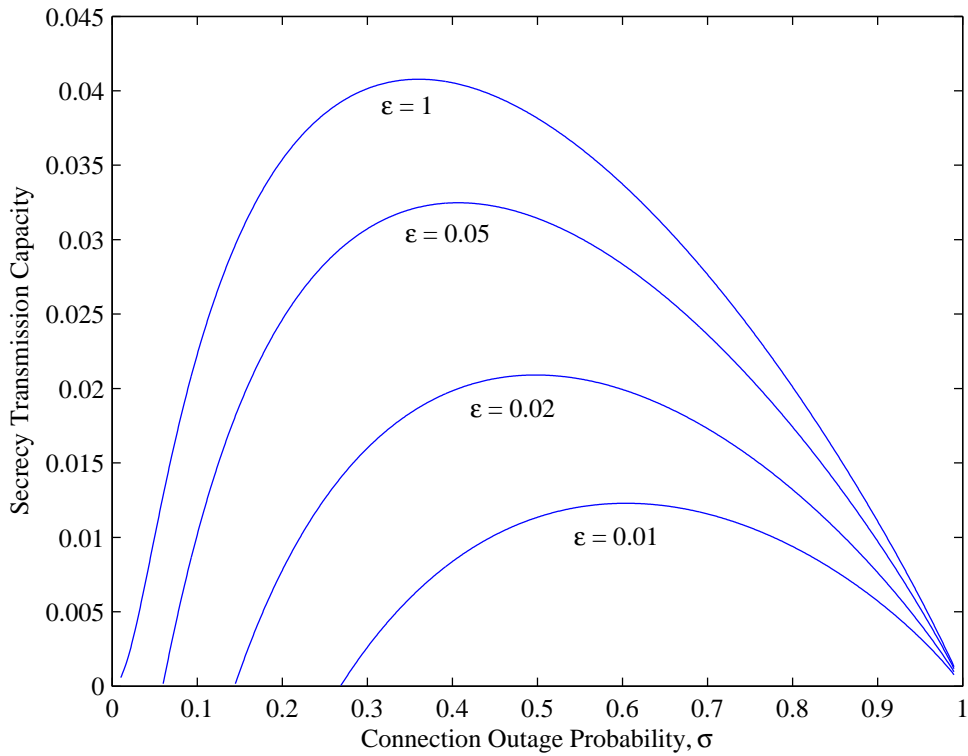}
\vspace{-0mm} \caption{The secrecy transmission capacity $\tau^{\rm{LB}}(r)$ in (\ref{eq:STCR}) versus the connection outage probability $\sigma$. Results are shown for networks with different secrecy outage constraints, \ie $\epsilon=0.01, 0.02, 0.05$, as well as no secrecy constraint, \ie $\epsilon=1$. The other system parameters are $r=1$, $\alpha=4$, $\lambda_l=0.01$, and $\lambda_e=0.001$.} \label{fig:STCvsSigma}
\end{figure}

Fig.~\ref{fig:STCvsSigma} shows the secrecy transmission capacity $\tau^{\rm{LB}}(r)$ in (\ref{eq:STCR}) versus the connection outage probability $\sigma$ with different security requirements. The feasible range of $\sigma$ for positive secrecy transmission capacity never reaches 0, which agrees with the result in (\ref{eq:OutRange}). We see that a moderate connection outage probability is desirable for achieving high secrecy transmission capacity. Furthermore, the optimal value of $\sigma$ increases as $\epsilon$ reduces. This is because that a larger $R_e$ is needed for a stronger security requirement, in which case larger $R_t$ and (hence) $\sigma$ are desirable for maximizing the secrecy transmission capacity. For example, the optimal $\sigma$ is 0.4 for $\epsilon=0.05$ while it increases to 0.5 for $\epsilon=0.02$ and to 0.6 for $\epsilon=0.01$. Note that the optimal value of $\sigma$ can be accurately computed from the closed-form expression in (\ref{eq:OptSig}) for sparse networks.

\begin{figure}[!t]
\centering\vspace{-0mm}
\includegraphics[width=0.85\columnwidth]{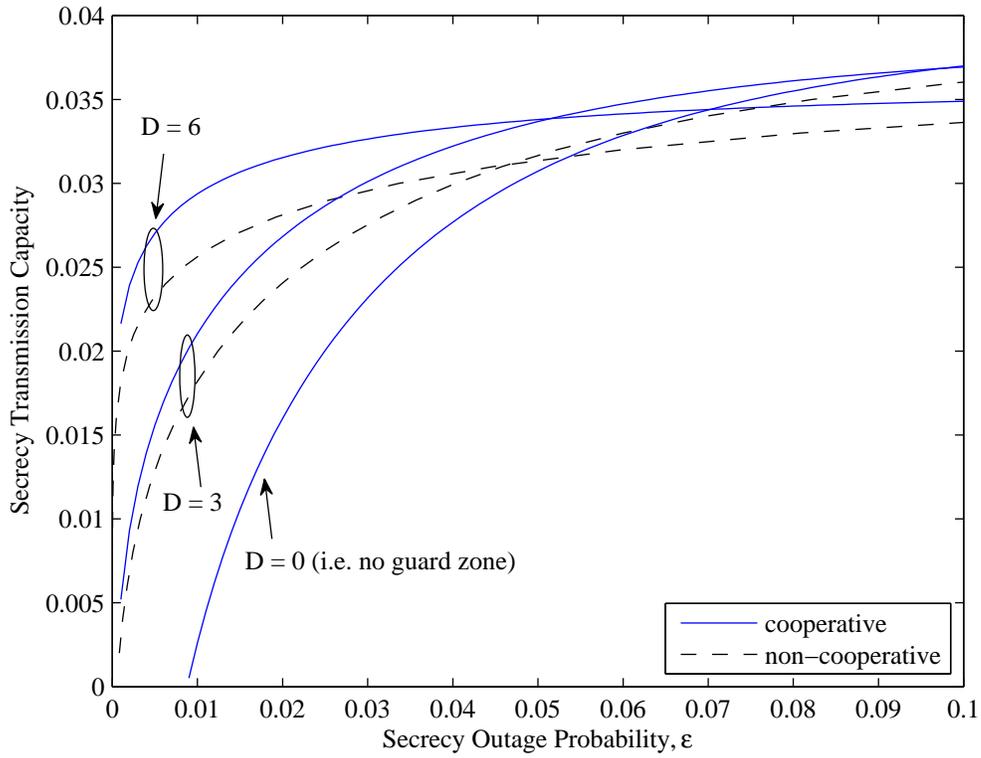}
\vspace{-0mm} \caption{The secrecy transmission capacity $\tau^{\rm{LB}}(r)$ with guard zone in (\ref{eq:STCRG0}) and (\ref{eq:STCRG}) versus the secrecy outage probability $\epsilon$. Results are shown for networks with different guard zone radii, \ie $D=0, 3, 6$. The other system parameters are $r=1$, $\alpha=4$, $\sigma=0.3$, $\lambda_l=0.01$, and $\lambda_e=0.001$.} \label{fig:STCvsSecZone}
\end{figure}

Now, we present the results on the use of a guard zone for improving the secrecy transmission capacity. Fig.~\ref{fig:STCvsSecZone} shows $\tau^{\rm{LB}}(r)$ in (\ref{eq:STCRG0}) and (\ref{eq:STCRG}) versus the secrecy outage probability $\epsilon$ for both the non-cooperative and cooperative protocols.\footnote{Although we have chosen $r=1$ in Fig.~\ref{fig:STCvsSecZone}, the benefit of using a secrecy guard zone demonstrated in Fig.~\ref{fig:STCvsSecZone} is also observed for other values of $r$ which can be either less than or greater than $D$ (plots omitted for brevity).} This figure clearly shows the remarkable benefit of guard zone for networks with high security requirements. For example, the secrecy transmission capacity at $\epsilon=0.01$ increases from 0.003 at $D=0$ (\ie no guard zone) to 0.018 with non-cooperative protocol and 0.021 with cooperative protocol at $D=3$. On the other hand, the benefit of guard zone reduces as the security requirement reduces. We also see that the cooperative protocol outperforms the non-cooperative protocol and the difference in secrecy transmission capacity is significant for networks with high security requirement. For example, the increase in $\tau^{\rm{LB}}(r)$ from the non-cooperative case to the cooperative case at $\epsilon=0.01$ is 17\% when $D=3$ and 14\% when $D=6$.

\begin{figure}[!t]
\centering\vspace{-0mm}
\includegraphics[width=0.85\columnwidth]{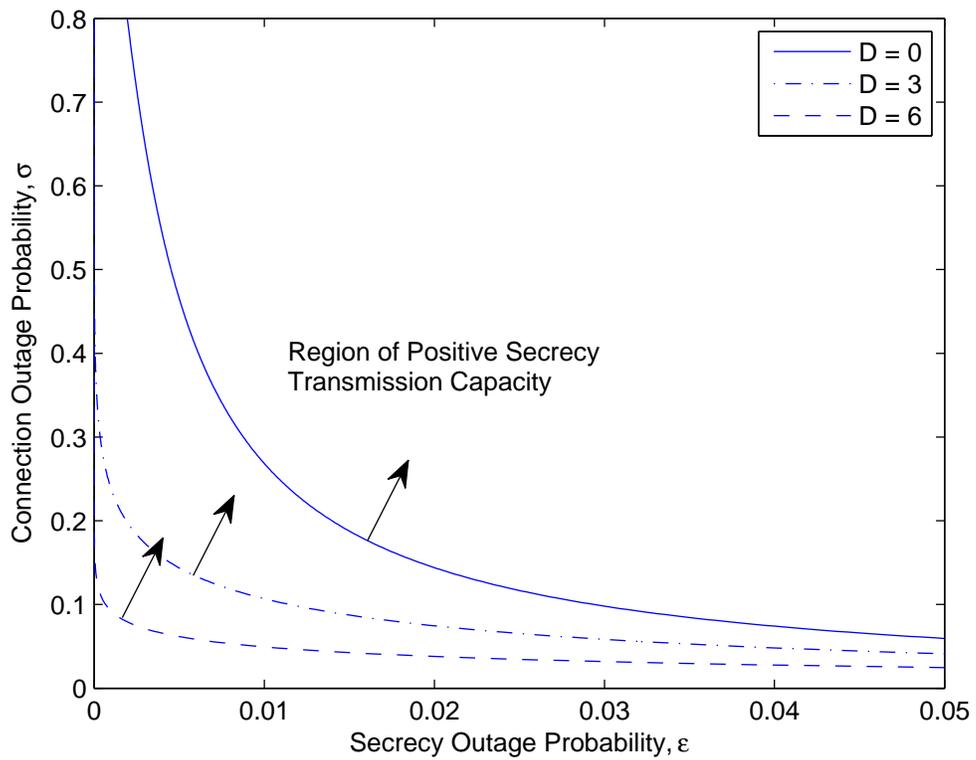}
\vspace{-0mm} \caption{The region of positive secrecy transmission capacity with and without secrecy guard zone. The case of cooperative transmitters is shown with different guard zone radii, \ie $D=0, 3, 6$. The curves are plotted based on the relationship between the connection outage probability $\sigma$ and the secrecy outage probability $\epsilon$ given in (\ref{eq:ReqG}). The other system parameters are $r=1$ and $\lambda_e=0.001$.} \label{fig:STCcondition}
\end{figure}

The feasible regions of the connection outage probability $\sigma$ and the secrecy outage probability $\epsilon$ for positive secrecy transmission capacity are illustrated in Fig.~\ref{fig:STCcondition}. The case of cooperative transmitters is considered when the guard zone is used. In general, it is impossible to have arbitrarily low outage probabilities while still operating at some positive secrecy transmission capacity. Nevertheless, the use of a guard zone greatly enlarges the feasible ranges of both outage probabilities.

\begin{figure}[!t]
\centering\vspace{-0mm}
\includegraphics[width=0.85\columnwidth]{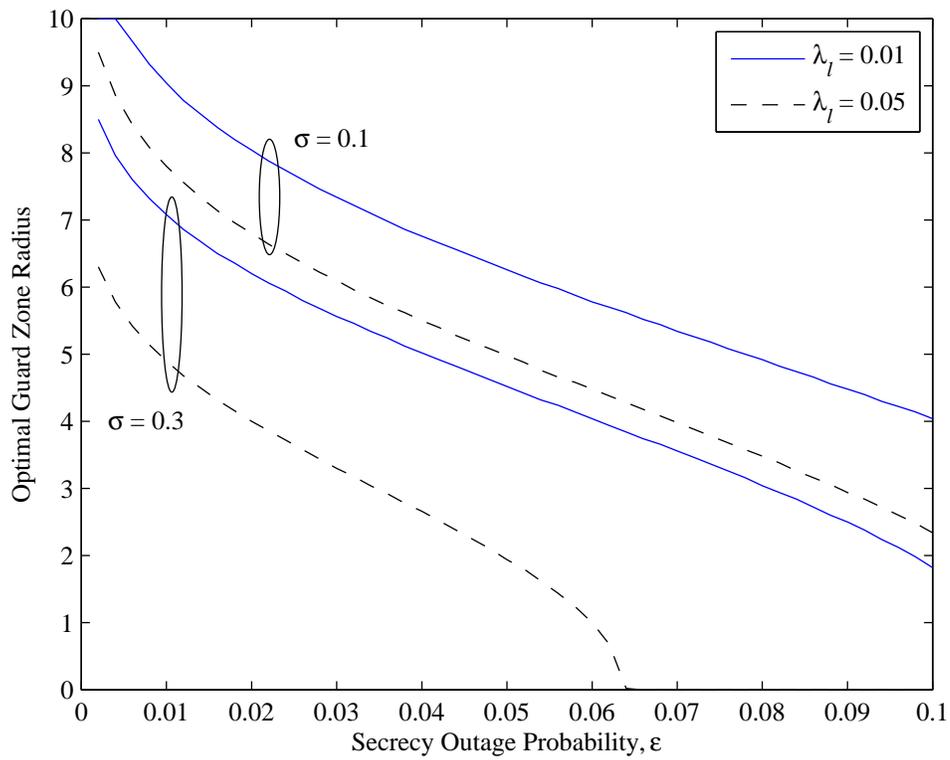}
\vspace{-0mm} \caption{The optimal secrecy guard zone radius versus the secrecy outage probability $\epsilon$ for the case of cooperative transmitters. Results are shown for networks with different connection outage constraints, \ie $\sigma=0.1, 0.3$, and different densities of legitimate transmitters, \ie $\lambda_l=0.01, 0.05$. The other system parameters are $r=1$, $\alpha=4$, and $\lambda_e=0.001$.} \label{fig:OptZonevsSec_AN}
\end{figure}

Fig.~\ref{fig:OptZonevsSec_AN} shows the optimal guard zone radius for cooperative transmitters. We see that the optimal value of $D$ reduces as the acceptable secrecy outage probability $\epsilon$ increases and it reaches zero at moderate to high values of $\epsilon$. We can also see the dependence of the optimal $D$ on $\sigma$ and $\lambda_l$. The general trend is that the optimal $D$ reduces as $\sigma$ or $\lambda_l$ increases. The dependence of the optimal $D$ on $\sigma$ agrees with our earlier observation from (\ref{eq:ReqG}). The dependence on $\lambda_l$ can be understood as follows: As the interference level (\ie $\lambda_l$) increases, the signal power received at the eavesdroppers is allowed to be higher for maintaining the same SIR. This in turn allows us to reduce the guard zone radius, which increases the spatial intensity of message transmissions. In practice, a legitimate transmitter may only be able to detect the existence of eavesdroppers in its vicinity, hence, the guard zone is usually small. Nevertheless, we have seen from Fig.~\ref{fig:STCvsSecZone} that a significant improvement in the secrecy transmission capacity can be achieved even with a small guard zone. Furthermore, by allowing a moderate connection outage probability, it is possible for the network to operate with the optimal guard zone radius which is within the maximum detection range of the transmitters.

\section{Conclusions}\label{sec:Conc}

In this work, we defined a performance metric named the secrecy transmission capacity, which was used to study the impact of physical layer security requirements on the throughput of large-scale decentralized wireless networks. Using tools and existing results from stochastic geometry, the secrecy transmission capacity can usually be characterized in simple analytical forms, as shown in this paper for Rayleigh fading channels. One important finding is that the throughput cost of achieving a moderate security level is relatively low, while it becomes very expensive to realize a highly secure network. In addition, we showed that the application of secrecy guard zone with artificial noise is a simple technique that can be used to dramatically reduce the throughput cost of achieving highly secure networks.

This model of secrecy transmission capacity can be extended to analyze and design networks with other transmission techniques, medium access control protocols, and eavesdropping strategies in future work. Similar to other transmission capacity formulations, the main limitation of this model is that it only considers single-hop transmissions, while the communication between an arbitrary source-destination pair usually requires multiple hops. End-to-end throughput analysis of wireless networks with physical layer security requirements is still an open problem. Another limitation of the current model is the homogeneous Poisson distribution of nodes. The impact of eavesdropper distribution on secrecy throughput is an interesting problem to investigate.

\bibliographystyle{IEEEtran}
\bibliography{IEEEabrv,journal_v5}

\newpage

\begin{table}[!t]
\caption{List of Notation} \centering\vspace{-3mm}
\begin{tabular}{|l|l|} \hline
     $\Phi_l$ & Poisson point process (PPP) of legitimate transmitter locations \\
     $\Phi_e$ & PPP of eavesdropper locations \\
     $\lambda_l$ & Density of $\Phi_l$ \\
     $\lambda_e$ & Density of $\Phi_e$ \\
     $R_t$ & Rate of the transmitted codewords \\
     $R_s$ & Rate of the confidential messages \\
     $R_e$ & $R_t-R_s$, rate loss for securing the messages against eavesdropping\\
     $\Pout$ & Connection outage probability \\
     $\Pint$ & Secrecy outage probability \\
     $\sigma$ & Constraint on $\Pout$ \\
     $\epsilon$ & Constraint on $\Pint$ \\
     $\tau(r)$ & Secrecy transmission capacity with transmission distance $r$\\
     $\beta_t$ & Threshold signal to interference ratio (SIR) for connection outage\\
     $\beta_e$ & Threshold SIR for secrecy outage\\
     $S$ & Rayleigh fading gain of the wireless channel\\
     $D$ & Radius of secrecy guard zones\\
     $\mathbb{P}(.)$ & Probability operator\\
     $\mathbb{E}\{.\}$ & Expectation operator\\\hline
\end{tabular}
\label{table:notations}
\vspace{9mm}
\end{table}

\newpage

\newpage

\newpage

\newpage

\newpage

\newpage

\end{document}